\documentclass{sig-alternate}
\usepackage[ruled,vlined]{algorithm2e}
\usepackage{graphicx}
\usepackage{url}
\usepackage{paralist}
\usepackage{tabularx}
\usepackage{balance}
\usepackage{multirow}
\usepackage{multicol}
\usepackage{pbox}
\usepackage{mathtools}
\usepackage[TABBOTCAP]{subfigure}
\usepackage{algorithmic}
\usepackage{graphicx}
\usepackage{paralist}
\usepackage{tabularx}
\usepackage{url}
\usepackage{balance}
\usepackage{multirow}
\usepackage{multicol}
\usepackage{amsmath}
\usepackage{hyperref}
\usepackage{verbatim}
\usepackage{float}
\usepackage[normalem]{ulem}
\usepackage{xspace}
\usepackage[usenames]{color}
\usepackage{colortbl}
\usepackage{amssymb}
\usepackage{ifthen}
\usepackage{pifont}
\usepackage{tikz}
\newcommand*\circled[1]{\tikz[baseline=(char.base)]{
            \node[shape=circle,draw,inner sep=0.5pt] (char) {#1};}}

\newcommand{\fusion}{{\sc Fusion}}
\newcommand{\fusions}{{\sc Fusion's}}

\include{macro}

\begin{document}
\toappear{}
\pagenumbering{gobble}
\title{Enhancing Android Application Bug Reporting}

\numberofauthors{1}
\author{
\alignauthor
Kevin Moran\\
\affaddr{College of William \& Mary}\\
\affaddr{Department of Computer Science}\\
\affaddr{P.O. Box 8795}\\
\affaddr{Williamsburg, VA 23187-8795, USA}\\
\email{kpmoran@cs.wm.edu}
}

\maketitle
\begin{abstract}
The modern software development landscape has seen a shift in focus toward mobile applications as smartphones and tablets near ubiquitous adoption. Due to this trend, the complexity of these ``apps" has been increasing, making development and maintenance challenging.  Current bug tracking systems do not effectively facilitate the creation of bug reports with useful information that will directly lead to a bug's resolution. To address the need for an improved reporting system, we introduce a novel solution, called \fusion, that helps reporters auto-complete reproduction steps in bug reports for mobile apps by taking advantage of their GUI-centric nature.  \fusion~links information, that reporters provide, to program artifacts extracted through static and dynamic analysis performed beforehand. This allows our system to facilitate the reporting process for developers and testers, while generating more reproducible bug reports with immediately actionable information.
\end{abstract}

\category{D.2.7}{Software Engineering}{Distribution, Maintenance, and Enhancement}

\terms{Experimentation, Design}

\keywords{Bug reports, android, reproduction steps, auto-completion} 

\section{Motivation \& Research Problem}
\label{sec:intro}
Software maintenance activities are known to be generally expensive and challenging \cite{25Tassey:NIST} and one of the most important maintenance tasks, especially in hyper-competitive mobile marketplaces, is bug report resolution.  However, current bug tracking systems such as Bugzilla \cite{bugzilla}, Mantis \cite{mantis}, the Google Code Issue Tracker (GCIT) \cite{google-code}, the GitHub Issue Tracker \cite{github-it}, and commercial solutions such as JIRA \cite{jira} rely mostly on unstructured natural language bug descriptions.  
	
The problem facing many of these current reporting systems is that typical natural language reports capture a coarse grained level of detail that makes developer reasoning about defects difficult.  This highlights the underlying \textit{\textbf{task}} that issue tracking systems must accomplish: \textit{bridging the lexical knowledge gap between typical reporters of a bug and the developers that must resolve the bugs.}    Previous studies on bug report quality and developer information needs highlight several factors that can impact the quality of reports \cite{15Breu:CSCW10, 4Joorabchi:MSR14, 3Bettenburg:FSE08}:  {\bf 1)} Other than ``Interbug dependencies'' (i.e., a situation where a bug was fixed in a previous patch), \textit{insufficient information} in bug reports is one of the leading causes of non-reproducible bug reports \cite{4Joorabchi:MSR14}; {\bf 2)} Developers consider (i)\textit{steps to reproduce}, (ii)\textit{stack traces}, and (iii)\textit{test cases/scenarios} as the most helpful sources of information in a bug report \cite{3Bettenburg:FSE08}; {\bf 3)} Information needs are greatest early in a bug's life cycle, therefore, a way to easily add the above features is important during bug report creation \cite{15Breu:CSCW10}.

Using these issues as motivation, we developed an augmented bug reporting solution, \fusion, that gleans contextual information from the GUI of a mobile app to accomplish the following goals: {\bf (i)} \textit{provide bug reports to developers with immediately actionable knowledge (reliable reproduction steps)} and {\bf (ii)} \textit{facilitate reporting by providing this information through an auto-completion mechanism.}

\section{Background \& Related Work}
\label{sec:related works}
\begin{figure*}[tb]
\centering
\vspace{-0.3cm}
\includegraphics[width=0.9\textwidth]{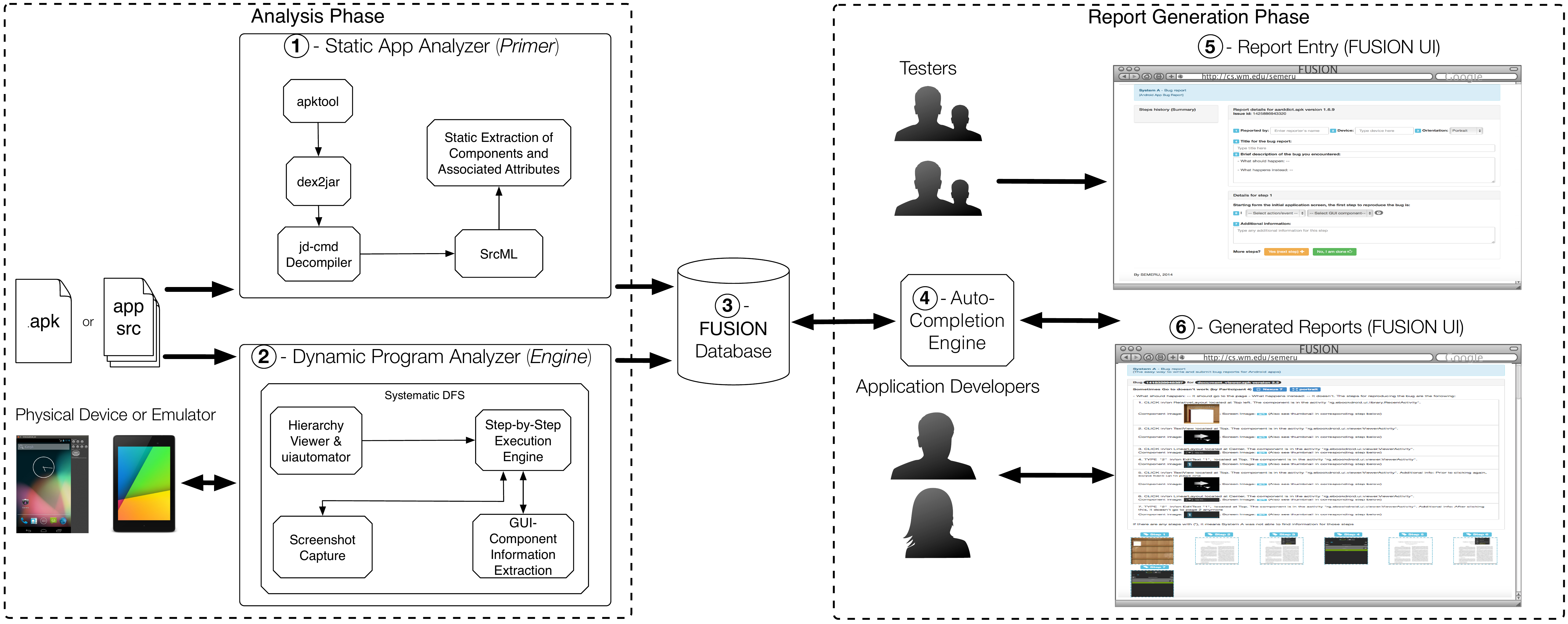}
\vspace{-0.4cm}
\caption{Overview of \fusion~Workflow}
\label{Design}
\vspace{-0.5cm}
\end{figure*}

\textit{\textbf{In-Field Failure Reproduction:}} A body of work known as in-field failure reproduction \cite{27Bell:ICSE13, 26Jin:ISSTA13, 8Zhou:ICSE12, 29Clause:ICSE07,18Jin:ICSE12,41Artzi:ECOOP2008,50Kifetew:ICST2014,43Cao:ASE14} shares similar goals with our approach. These techniques collect run-time information (e.g., execution traces) from instrumented programs that provide developers with a better understanding of the causes of a failure, which subsequently helps expedite the fixing of the corresponding faults.  However, there are three key points that set our work apart and illustrate how \fusion~improves on the state of research.   

	\textit{First}, in-field failure reproduction techniques rely on potentially expensive program instrumentation.  \fusion~is completely automatic, our static and dynamic analysis techniques only need to be applied once for the version of the program that is released for testing.  \textit{Second}, current in-field failure reproduction techniques require an oracle to signify when a failure has occurred (e.g., a crash).  \fusion~is not an approach for crash or failure detection, it is designed to support testers during the bug reporting process. \textit{Third}, these techniques have not been applied to mobile apps and would likely need to be optimized further to be applicable for the corresponding resource-constrained environment.  

\section{Approach}
\label{sec:approach}
    In this section we present our approach for Android bug reporting that utilizes an \textit{Analyze} $\rightarrow$ \textit{Generate} workflow corresponding to two major phases. In the \texttt{Analysis Phase} \fusion~collects app and GUI information and in the \texttt{Report Generation Phase}, \fusion~operates on this information to aid users in constructing bug reports.  The overall design of \fusion~can be seen in Figure \ref{Design}.  Please refer to the following materials \cite{Moran:FSE15, appendix} for a complete description and dataset.

\subsection{Analysis Phase} 

The \textit{Analysis Phase} has two major components: 1) static analysis \emph{(Primer)}, and 2) dynamic program analysis \emph{(Engine)} of a target app.  Both components store their data in the \fusion~database (Fig. \ref{Design} - \circled{3}).
    
\textit{\textbf{Static Analysis (Primer)}} The goal of the \emph{Primer} (Fig. \ref{Design} - \circled{1})
   is to extract all of the GUI components and associated information from the app source code.   For each GUI component, the \emph{Primer} extracts: (i) possible actions on that component, (ii) type of the component (e.g., Button, Spinner), (iii) activities the component is contained within, and (iv) class files where the component is instantiated.  Thus, this phase results in a universe of possible GUI components for the application, and establishes traceability links connecting these components to specific code-level artifacts.  
    
\textit{\textbf{Dynamic Analysis (Engine)}}  The \emph{Engine} (Fig. \ref{Design} - \circled{2}) is used to glean dynamic contextual information, such as the location of the GUI component on the screen, and enhance the database with both run-time GUI and application event flow information. The goal of the \emph{Engine} is to explore an app in a systematic manner ripping and extracting run-time information related to the GUI components during execution including: (i) text associated with a component, (ii) contextual screenshots of the component, (iii) the activity or screen the component is located within, and (iv) the location of the component on a given app screen.  During the ripping, before each step is executed on the GUI, the \emph{Engine} calls {\tt UIAutomator} subroutines to extract the contextual information outlined above regarding each GUI component on the screen.  To effectively explore and model the application we took inspiration from our previous work \cite{Linares:MSR15} and implemented a systematic depth-first search (DFS) algorithm that performs click/tap events on the components in the GUI hierarchy for a given app.

\subsection{Report Generation Phase}  
         
         The design of the \textit{Report Generation Phase} component of \fusion~has two goals: {\bf 1)} Allow for traditional natural language input in order to give a high-level overview of a bug. {\bf 2)} Auto-complete the reproduction steps of a bug through suggestions derived by tracking the position of the reporter's step entry in the app event flow. During the \emph{Report Generation Phase}  \fusions~Auto-completion engine populates a decision tree based on the extracted app model and aids the reporter in constructing the steps needed to recreate a bug by making suggestions based upon the ``potential" GUI state reached by the declared steps.  This means for each step $s$,  \fusion~infers -- online -- the GUI state $GUI_s$ in which the target app should be by taking into account the step history.   For each step,  \fusion~verifies that the suggestion made to the reporter is correct by asking the reporter to select a full contextual screenshot corresponding to the \texttt{\{action ,gui-component\}} tuple they intended to record.  The end result is a detailed bug report including natural-language general information and contextualized reproduction steps with screen-shots for each step and traceability links from steps to the app source code.

\section{Results and Contributions}
\label{sec:results}
We have evaluated the proposed framework with two empirical studies using 15 real bugs in 14 open-source Android applications extracted from the F-Droid marketplace \cite{fdroid} comparing \fusion~against the GCIT \cite{google-code} and the original bug reports in two maintenance activities involving reporting (\textit{Study 1}) and reproducing bugs (\textit{Study 2)}.  \textit{Study 1} involved eight graduate students from the College of William \& Mary (W\&M) who created reports from videos of the bugs for the two systems, for a total of 120 (60 \fusion, 60 GCIT) bug reports.  In \textit{Study 2}, 20 new participants, also W\&M students, attempted to reproduce the 15 bugs from the bug reports created in the first study, along with the original issue tracker reports.  Our results show that, overall \fusion~produces more reproducible bug reports than current issue trackers (13/120 non-reproducible with \fusion, versus 23/120 non-reproducible with GCIT) . The major contribution of this work is the following conclusion: \textit{By leveraging program analysis and the GUI-centric nature of mobile apps, mobile issue trackers can be augmented to facilitate users creating more detailed and reproducible bug reports.}

\balance

\bibliography{main}

\begin{thebibliography}{10}

\bibitem{bugzilla}
Bugzilla issue tracker \url{https://bugzilla.mozilla.org}.

\bibitem{fdroid}
F-droid. \url{https://f-droid.org/}.

\bibitem{github-it}
Github issue tracker \url{https://github.com/features}.

\bibitem{google-code}
Google code issue tracker
  \url{https://code.google.com/p/support/wiki/IssueTracker}.

\bibitem{jira}
Jira bug reporting system \url{https://www.atlassian.com/software/jira}.

\bibitem{mantis}
Mantis bug reporting system \url{https://www.mantisbt.org}.

\bibitem{41Artzi:ECOOP2008}
S.~Artzi, S.~Kim, and M.~Ernst.
\newblock Recrash: Making software failures reproducible by preserving object
  states.
\newblock In J.~Vitek, editor, {\em ECOOP 2008 -- Object-Oriented Programming},
  volume 5142 of {\em Lecture Notes in Computer Science}, pages 542--565.
  Springer Berlin Heidelberg, 2008.

\bibitem{27Bell:ICSE13}
J.~Bell, N.~Sarda, and G.~Kaiser.
\newblock Chronicler: Lightweight recording to reproduce field failures.
\newblock In {\em Proceedings of the 2013 International Conference on Software
  Engineering}, ICSE '13, pages 362--371, Piscataway, NJ, USA, 2013. IEEE
  Press.

\bibitem{3Bettenburg:FSE08}
N.~Bettenburg, S.~Just, A.~Schr\"{o}ter, C.~Weiss, R.~Premraj, and
  T.~Zimmermann.
\newblock What makes a good bug report?
\newblock In {\em Proceedings of the 16th ACM SIGSOFT International Symposium
  on Foundations of Software Engineering}, SIGSOFT '08/FSE-16, pages 308--318,
  New York, NY, USA, 2008. ACM.

\bibitem{15Breu:CSCW10}
S.~Breu, R.~Premraj, J.~Sillito, and T.~Zimmermann.
\newblock Information needs in bug reports: Improving cooperation between
  developers and users.
\newblock In {\em Proceedings of the 2010 ACM Conference on Computer Supported
  Cooperative Work}, CSCW '10, pages 301--310, New York, NY, USA, 2010. ACM.

\bibitem{43Cao:ASE14}
Y.~Cao, H.~Zhang, and S.~Ding.
\newblock Symcrash: Selective recording for reproducing crashes.
\newblock In {\em Proceedings of the 29th ACM/IEEE International Conference on
  Automated Software Engineering}, ASE '14, pages 791--802, New York, NY, USA,
  2014. ACM.

\bibitem{29Clause:ICSE07}
J.~Clause and A.~Orso.
\newblock A technique for enabling and supporting debugging of field failures.
\newblock In {\em Proceedings of the 29th International Conference on Software
  Engineering}, ICSE '07, pages 261--270, Washington, DC, USA, 2007. IEEE
  Computer Society.

\bibitem{4Joorabchi:MSR14}
M.~Erfani~Joorabchi, M.~Mirzaaghaei, and A.~Mesbah.
\newblock Works for me! characterizing non-reproducible bug reports.
\newblock In {\em Proceedings of the 11th Working Conference on Mining Software
  Repositories}, MSR 2014, pages 62--71, New York, NY, USA, 2014. ACM.

\bibitem{18Jin:ICSE12}
W.~Jin and A.~Orso.
\newblock Bugredux: Reproducing field failures for in-house debugging.
\newblock In {\em Proceedings of the 34th International Conference on Software
  Engineering}, ICSE '12, pages 474--484, Piscataway, NJ, USA, 2012. IEEE
  Press.

\bibitem{26Jin:ISSTA13}
W.~Jin and A.~Orso.
\newblock F3: Fault localization for field failures.
\newblock In {\em Proceedings of the 2013 International Symposium on Software
  Testing and Analysis}, ISSTA 2013, pages 213--223, New York, NY, USA, 2013.
  ACM.

\bibitem{50Kifetew:ICST2014}
F.~Kifetew, W.~Jin, R.~Tiella, A.~Orso, and P.~Tonella.
\newblock Reproducing field failures for programs with complex grammar-based
  input.
\newblock In {\em Software Testing, Verification and Validation (ICST), 2014
  IEEE Seventh International Conference on}, pages 163--172, March 2014.

\bibitem{Linares:MSR15}
M.~Linares-V\'{a}squez, M.~White, C.~Bernal-C\'{a}rdenas, K.~Moran, and
  D.~Poshyvanyk.
\newblock Mining android app usages for generating actionable gui-based
  execution scenarios.
\newblock In {\em 12th Working Conference on Mining Software Repositories
  (MSR'15)}, to appear, 2015.

\bibitem{Moran:FSE15}
K.~Moran, M.~Linares-V\'{a}squez, C.~Bernal-C\'{a}rdenas, and D.~Poshyvanyk.
\newblock Auto-completing bug reports for android applications.
\newblock In {\em 10th Joint Meeting of the European Software Engineering
  Conference and the ACM SIGSOFT Symposium on the Foundations of Software
  Engineering (ESEC/FSE'15)}, to appear, 2015.

\bibitem{appendix}
K.~Moran, M.~L. Vasquez, C.~B. Cardenas, and D.~Poshyvanyk.
\newblock Fusion online appendix \url{http://www.fusion-android.com}.

\bibitem{25Tassey:NIST}
G.~Tassey.
\newblock The economic impacts of inadequate infrastructure for software
  testing.
\newblock Technical report, National Institute of Standards and Technology,
  2002.

\bibitem{8Zhou:ICSE12}
J.~Zhou, H.~Zhang, and D.~Lo.
\newblock Where should the bugs be fixed? - more accurate information
  retrieval-based bug localization based on bug reports.
\newblock In {\em Proceedings of the 34th International Conference on Software
  Engineering}, ICSE '12, pages 14--24, Piscataway, NJ, USA, 2012. IEEE Press.

\end{thebibliography}
\bibliographystyle{abbrv}
\balancecolumns

\end{document}